\documentclass[a4paper,11pt]{article}
\usepackage{pos}

\usepackage{multirow}
\usepackage{rotating}
\usepackage{mathrsfs}
\usepackage{booktabs}
\usepackage{longtable}
\usepackage{placeins}
\usepackage{upgreek}
\usepackage{svg}
\usepackage{comment}
\usepackage{pdfpages}
\usepackage{siunitx}
\usepackage{lineno}
\usepackage[export]{adjustbox}

\title{Contribution title}

\author[a]{M.~Davis}
\author[a]{G.~Stage}
\author[a]{A.~Borjigin}
\author[a]{S.~Beringer}
\author[a]{N.~Lynch}
\author[a]{S.~Nakarmi}
\author[a]{C.~Galmes Altafulla}
\author[a]{A.~Drumm}
\author[a]{A.~Molnar}
\author[a]{A.~Tiernan}
\author*[a]{S.~M.~Mazza}
\author[a]{H.~Sadrozinski}
\author[a]{B.~A.~Schumm}
\author[a]{A.~Seiden}
\author[a]{F.~MartinezMckinney}
\author[a]{T.~Shin}

\affiliation[a]{SCIPP, UC Santa Cruz,\\
  1156 High St, Santa Cruz (CA), U.S.}

\emailAdd{simazza@ucsc.edu}

\abstract{
Low-Gain Avalanche Detectors (LGADs) are characterized by a fast rise time (500~ps) and extremely good time resolution (down to 17~ps). 
The intrinsic low granularity of LGADs and the large power consumption of readout chips for precise timing are problematic in near-future experiments such as e+e- Higgs factories (FCC-ee) and the ePIC detector at the Electron-Ion Collider. 
AC-coupled LGADs, where the readout metal is AC-coupled through an insulating oxide layer, could solve both issues at the same time thanks to the 100\% fill factor and charge-sharing capabilities. Charge sharing between electrodes allows a hit position resolution well below the pitch/$\sqrt12$ of standard segmented detectors. At the same time, it relaxes the channel density and power consumption requirement of readout chips. Extensive laboratory characterization of AC-LGAD devices from the first full-size (up to 3x4 cm) production from HPK for ePIC will be shown in this contribution. Both pixel and strip geometry was produced and tested. 
This study was conducted within the scope of the ePIC detector time of flight (TOF) layer R\&D program at the EIC.
}

\FullConference{%
  VERTEX2025\\
  Knoxville, USA}

\usepackage{cleveref}
\title{Characterization of the first full-size production for ePIC TOF layers}
\date{October 2025}

\begin{document}

\maketitle

\section{Introduction - 6 pages excluding abstract and references}

\begin{figure}[!hbt]
 \centering
 \includegraphics[width=0.3\columnwidth]{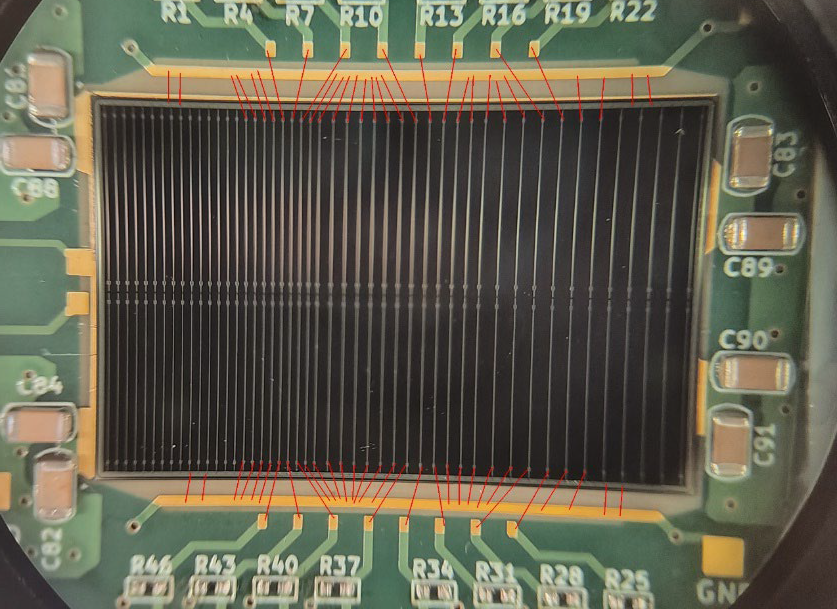}
 \includegraphics[width=0.2\columnwidth]{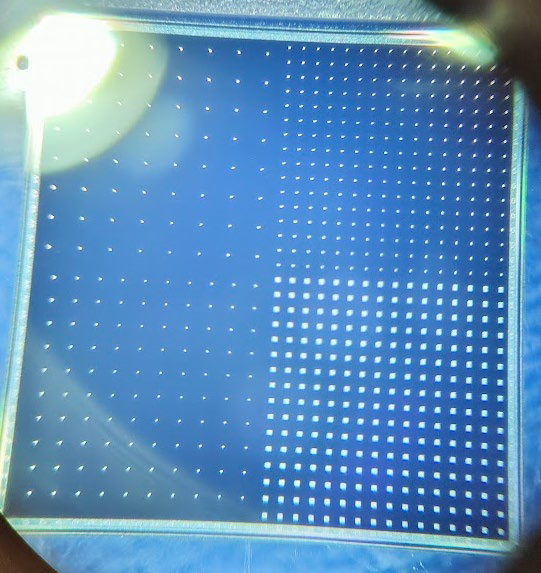}
 \includegraphics[width=0.3\columnwidth]{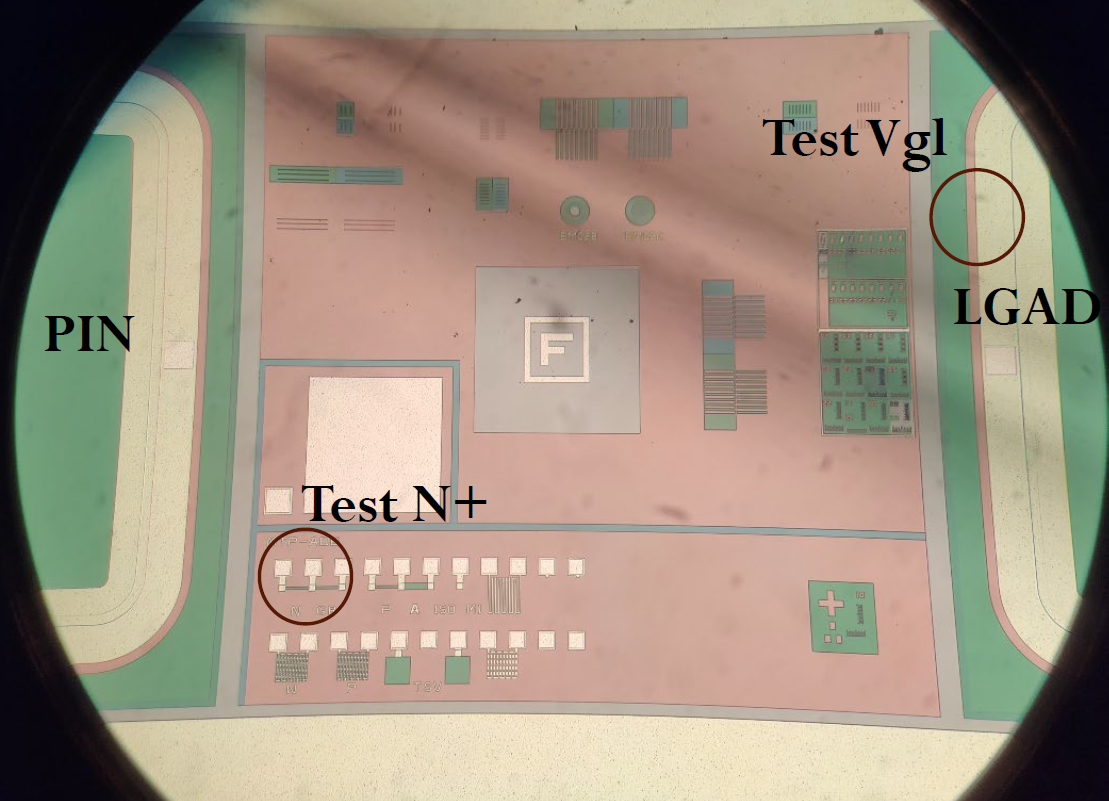}
  \caption{Picture of a full-size strip sensor (Left), full-size pixel sensor (center) and test structures (Right).}
 \label{fig:ACLGAD}
\end{figure}

Low Gain Avalanche Detectors (LGADs) have been established over the past decade as a leading fast-timing silicon sensor technology~\cite{Pellegrini:2014lki, Sadrozinski:2013nja}, achieving time resolutions on the order of tens of picoseconds.
In their first large-scale applications, the High Granularity Timing Detector (HGTD) in ATLAS~\cite{Mazza:2019dkn} and the MIP Timing Detector (MTD) in CMS~\cite{Ferrero:2022ynt}, the segmentation is limited to pad sizes of approximately 1 mm pitch, constrained by considerations of power consumption, fill factor, and electric field uniformity.
These limitations are addressed in the AC-LGAD technology, also known as the Resistive Silicon Detector (RSD)~\cite{BISHOP2024169478,Mandurrino}.
In this design, four layers of the sensor, the P-type bulk, P$\textsuperscript{++}$ gain layer, N$\textsuperscript{+}$ layer, and a dielectric insulator, are integrated into common sheets, separating the active region from the segmented metal readout electrodes.
Signals generated in the bulk and amplified in the gain layer are capacitively coupled to multiple readout pads, enabling charge sharing among neighboring channels.
This configuration allows precise position reconstruction—achieving spatial resolutions of a few percent of the readout pitch ($<5\%$~\cite{MENZIO2024169526}) while maintaining manageable power density and excellent timing performance.
AC-LGADs have been selected as the sensor technology for the time-of-flight (TOF) system of the ePIC detector~\cite{AbdulKhalek:2021gbh}, which employs strip sensors in the barrel region and pixel sensors in the end-cap regions.
Throughout the lifetime of ePIC, the TOF layers are expected to experience fluences up to approximately $10^{13}$~$\mathrm{n_{eq}}/\mathrm{cm}^2$ in the most exposed regions.
The performance of irradiated small-scale AC-LGAD prototypes from Hamamatsu Photonics K.K. (HPK)~\cite{Hamamatsu} have been reported in~\cite{Stage:2025jre}.
The performance of the first full-size prototype sensors from HPK for the ePIC TOF layer will be presented in this paper. A full-size strip sensor and pixel sensor are shown in Fig.~\ref{fig:ACLGAD}.

\section{ePIC TOF sensor fabrication at HPK}
Several wafers of both pixel and strip devices were produced with different geometries and thicknesses as summarized in Tab.~\ref{tab:wafers}.
The barrel TOF nominal dimensions for the strips are 3x2~cm, with 1~cm long strips and a pitch of 500~$\mu$m divided into two segments. 3x1~cm sensors were produced as well with the same strip geometry, which in the final design will populate the edge of the modules. 
Some devices with 3x4~cm with four strip segments were produced; however, this geometry was for an outdated layout for the barrel TOF.
The end-cap TOF nominal dimension for the pixels is 1.6x1.6~cm with a pitch of 500~$\mu$m.
Even though the baseline pitch for the TOF is 500~$\mu$m, this production explored the possibility of a larger pitch to reduce the channel density and total count.
Previous productions~\cite{Stage:2025jre,BISHOP2024169478} provided the input for resistivity and dielectric thickness; for the new production, resistivity is E-type (1600~$\Omega/\square$) for strips and C-type (400~$\Omega/\square$) for pixel detectors, the dielectric capacitance is 600~$pF/mm^2$ for all wafers.

\begin{table}[ht]
\centering
\begin{tabular}{|l|lll||ll|} 
\hline
Geometry & Wafers & thickness ($\mu$m) & Dimension (cm) & size ($\mu$m) & pitch ($\mu$m)\\
\hline
Strips &W2,6,11,12 & 50 & 3x1, 3x2, 3x4 & 40 & 500\\
&W13,15,22,23 & 30 & 3x1, 3x2, 3x4 & 50 & 500\\
&&&& 50 & 750\\
&&&& 50 & 1000\\
\hline
Pixels & W7,8 & 30 & 1.6x1.6 & 150 & 500\\
& W3,4 & 20 & 1.6x1.6 & 50-100 & 500\\
&&&& 50-100 & 750\\
&&&& 50-100 & 1000\\
\hline
\end{tabular}
\caption {Geometry and thickness of the tested HPK AC-LGAD wafers.}
\label{tab:wafers}
\end{table}

\section{Experimental procedure}
The sensors were first characterized on a probe station using current–voltage (IV) and capacitance–voltage (CV) measurements to evaluate breakdown behavior and homogeneity of the gain layer. 
The resistivity was measured using test structures at the edge of the wafers.
Subsequently, the strip devices were studied with a focused-laser Transient Current Technique (TCT) setup to assess the charge-sharing behavior, signal rise time, and time of arrival.
The jitter component of the time resolution and the position jitter were estimated from the laser TCT measurements with the same technique detailed in~\cite{BISHOP2024169478}.

\subsection{CV/IV Probing Station}
The sensors were electrically characterized with a precision probe station to evaluate the production yield, the homogeneity of the gain layer, and the N+ resistivity.
The probes were connected to a HV power supply and an LCR meter to measure current, Inductance, capacitance, and resistance. 
HV is supplied from the backside of the sensor.
Measurements of current vs voltage (IV) were taken to analyze the breakdown of the sensor and production yield; additionally, IV measurements were taken to measure the N+ layer resistive homogeneity using the test structures.
Capacitance vs voltage (CV) scans were performed to measure the gain layer's depletion voltage, $\mathrm{V_{GL}}$, which is proportional to the gain layer doping.

\subsection{Laser TCT Station}
Charge collection measurements using the Transient Current Technique (TCT) follow the procedure described in~\cite{Ott:2022itj,BISHOP2024169478,Stage:2025jre}.
Sensors were mounted on 16-channel fast amplifier boards (1~GHz bandwidth) developed at Fermilab (FNAL)~\cite{Heller:2022aug} and read out with a 2~GHz, 20~Gs/s oscilloscope.
An infrared (1064~nm) pulsed laser\footnote{NKT photonics KATANA10 laser, out of production} with 30~ps pulse width and 10–20~$\mu$m spot size was used to simulate the response of a MIP in silicon.
Since the IR laser does not penetrate the metal, scans were performed only in the inter-metal regions.
The readout board was mounted on X–Y stages with $\sim$1~$\mu$m precision~\footnote{STANDA 8MT173, \url{https://www.standa.lt/products/catalog/motorised_positioners?item=59}},
allowing position-dependent response studies. Scans were carried out in 10-50~$\mu$m steps.
At each position, 100 waveforms were averaged to reduce laser power fluctuations, which were further corrected using a reference photodiode.
The pulse shape at each step was analyzed to extract the pulse maximum (Pmax), 10-90\% rise time, and time of arrival following the CFD50 method (time at the 50\% of the maximum).
The laser is set to inject roughly one minimum ionizing particle (MIP) worth of charge; however, since the laser power changes over time, the actual injected charge can differ by up to 50\%.

\section{Electrical characterization}

\subsection{Yield}
\begin{figure}[!hbt]
 \centering
 \includegraphics[width=0.8\columnwidth]{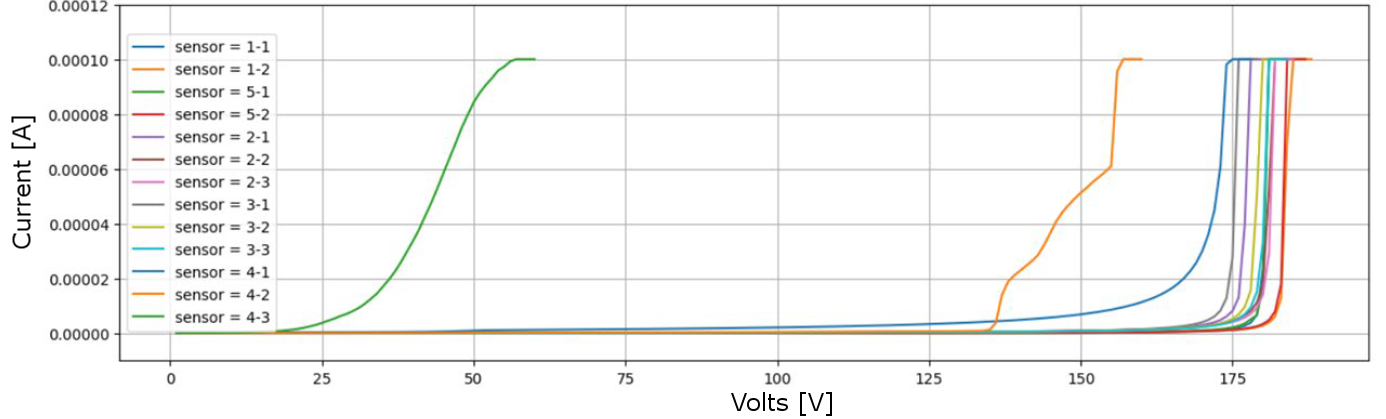}
  \caption{Current over voltage (IV) of one 30~$\mu$m wafer of strip detectors.}
 \label{fig:strips_IV}
\end{figure}

All devices were characterized with a current over voltage (IV) measurement. The breakdown voltage is evaluated as the voltage at which the devices's current increases exponentially to tens of $\mu$A. Fig.~\ref{fig:strips_IV} shows the IV of a full wafer of 30~$\mu$m strips; two devices show an early breakdown, and one device shows an increased current. 
The yield for the 30~$\mu$m strips is around 80\%, while the yield of the 50~$\mu$m strips is barely above 50\% due to issues in the production that were understood and corrected by HPK in the following production.
The yield for the pixel sensors is around 90\%.

\subsection{Gain Homogeneity}
\begin{figure}[h]
    \centering
    \includegraphics[width=0.35\linewidth]{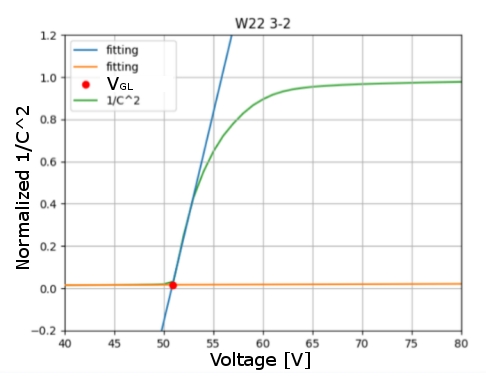}
    \includegraphics[width=0.42\linewidth]{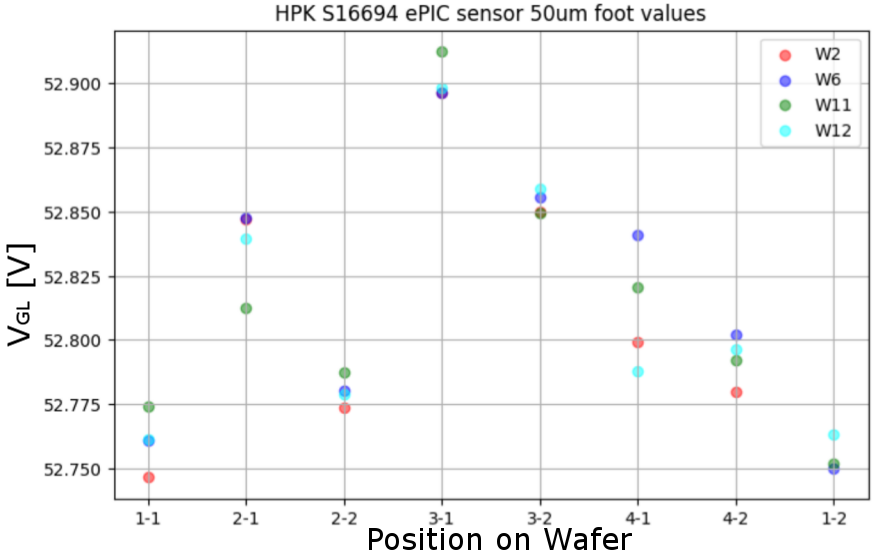}
    \caption{Left: Linear fittings of $\mathrm{1/C^2}$ to find the gain layer depletion $\mathrm{V_{GL}}$. Right: variation of $\mathrm{V_{GL}}$ across a wafer (X-Y) for 50~$\mu$m strip detectors. }
    \label{fig:gain_fits}
\end{figure}

The gain layer homogeneity for the fabricated wafers was measured as the gain layer depletion voltage ($\mathrm{V_{GL}}$) using the capacitance over voltage (CV) measurement.
The CV measurements were performed on conventional LGADs positioned along the wafer's perimeter (Fig.~\ref{fig:ACLGAD}, Right).
The $\mathrm{1/C^2}$ response as a function of bias voltage was used to evaluate $\mathrm{V_{GL}}$, which corresponds to the sharp variation at the depletion of the gain layer.
The resulting $\mathrm{1/C^2}$ plots were fitted with two linear fits, one corresponding to the bulk and the other to the gain layer, and the intersection point of these fits was extracted, as shown in Figure~\ref{fig:gain_fits}. 

\subsection{N+ resistivity}
The test structure in Fig~\ref{fig:ACLGAD} (Right) was used to measure the resistivity of the N+ layer. An IV is performed at the ends of the test structure and linearly fitted to measure the resistance, which is then divided by the number of squares (10) to calculate the resistivity. The resistivity for the strips wafers is around 2000~$\Omega$, while the one for the pixel wafers is around 550~$\Omega$, which are consistent with the expected values.

\textbf{\begin{table}[ht]
\centering
\begin{tabular}{|ll|llllll|} 
\hline
Geo. & Thick. & Yield (\%) & Breakdown (V) & $\mathrm{V_{GL}}$ (V) & $\sigma \mathrm{V_{GL}}$ (V) & N+ ($\Omega$) & $\sigma$N+ ($\Omega$)\\
\hline
Strips & 50~$\mu$m & 57 & 180 & 55.4 & 0.3 & 1981 & 109\\
& 30~$\mu$m & 80 & 180 & 50.7 & 0.2 & 2173 & 131\\
\hline
Pixels & 30~$\mu$m & 89 & 175 & 47.8 & 0.1 & 541 & 40\\
& 20~$\mu$m & 93 & 115 & 40.2 & 0.2 & 539 & 39\\
\hline
\end{tabular}
\caption {Summary of wafer measurements. $\sigma V_{GL}$ corresponds to the 1$\sigma$ standard deviation of $\mathrm{V_{GL}}$, similarly for  $\sigma$N+.}
\label{tab:wafers}
\end{table}
}

\section{Dynamic characterization}
The strip sensors were tested with the laser TCT system. 
The two thicknesses and all geometries were tested except for strips of 40~$\mu$m width and pitch of 500~$\mu$m.
The pixel sensors' dynamic characterization will be the topic of a future paper, as the devices are still under study.

\subsection{Signal characteristics}
\begin{figure}[h]
    \centering
    \includegraphics[width=0.35\linewidth]{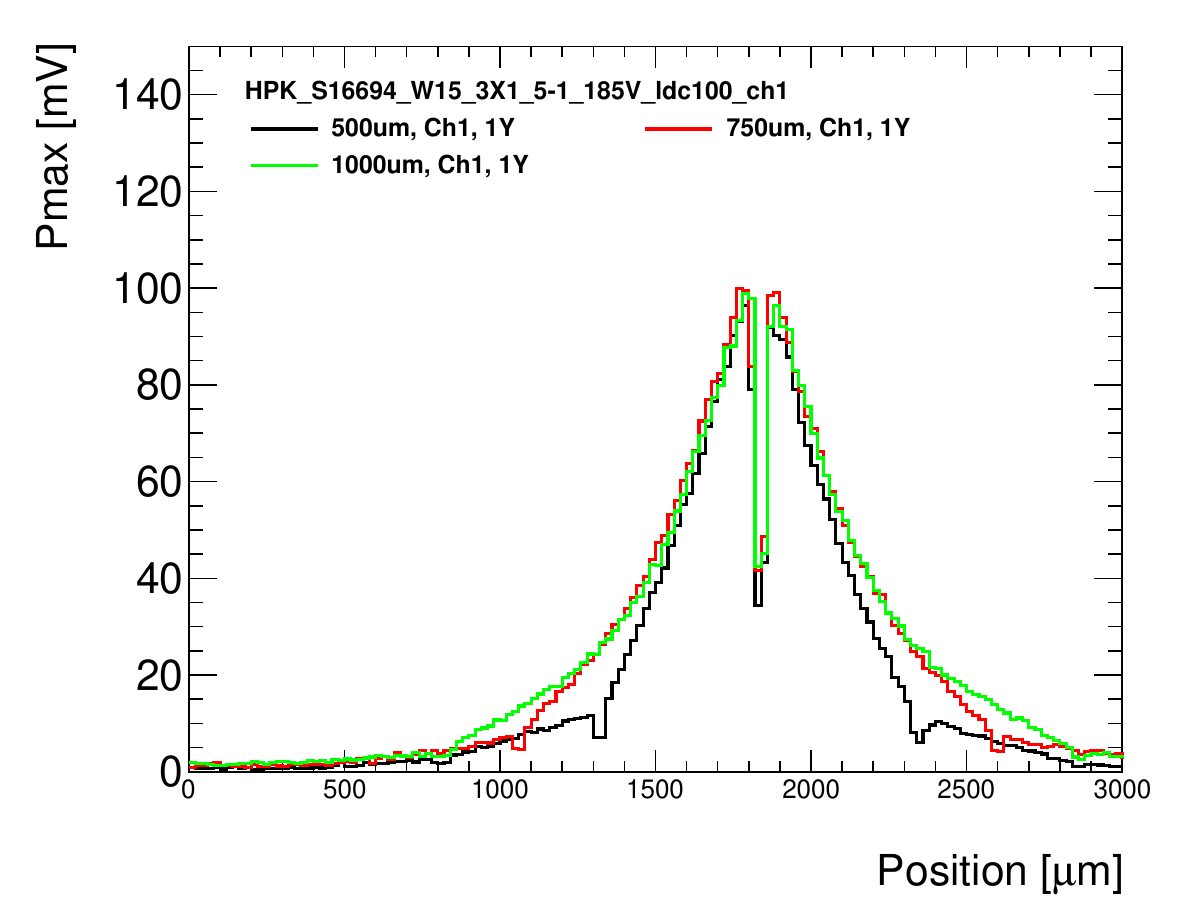}
    \includegraphics[width=0.35\linewidth]{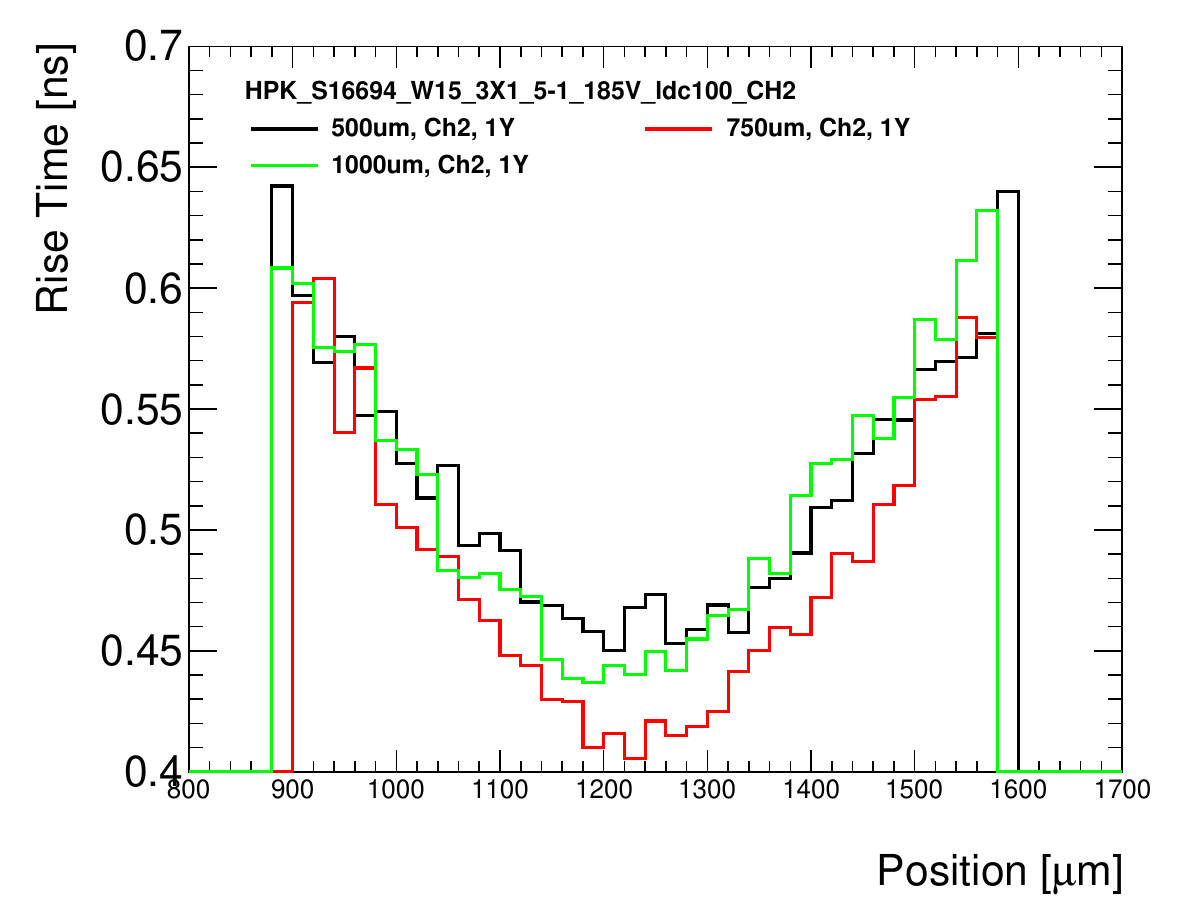}
    \caption{Left: Pmax as a function of laser injection position perpendicular to the strip. Right: rise time of the pulse vs laser position.}
    \label{fig:strip_pmax}
\end{figure}

\begin{figure}[h]
    \centering
    \includegraphics[width=0.35\linewidth,valign=t]{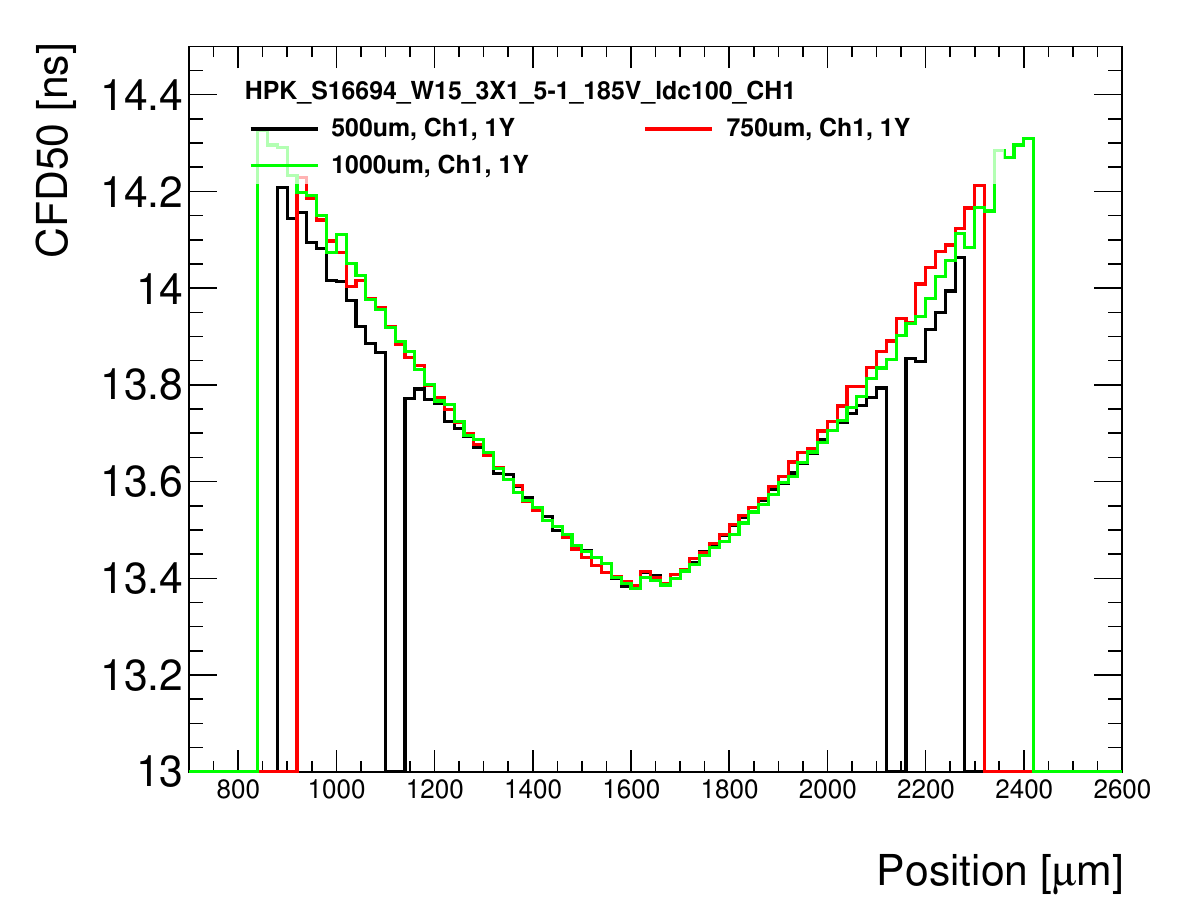}
    \includegraphics[width=0.36\linewidth,valign=t]{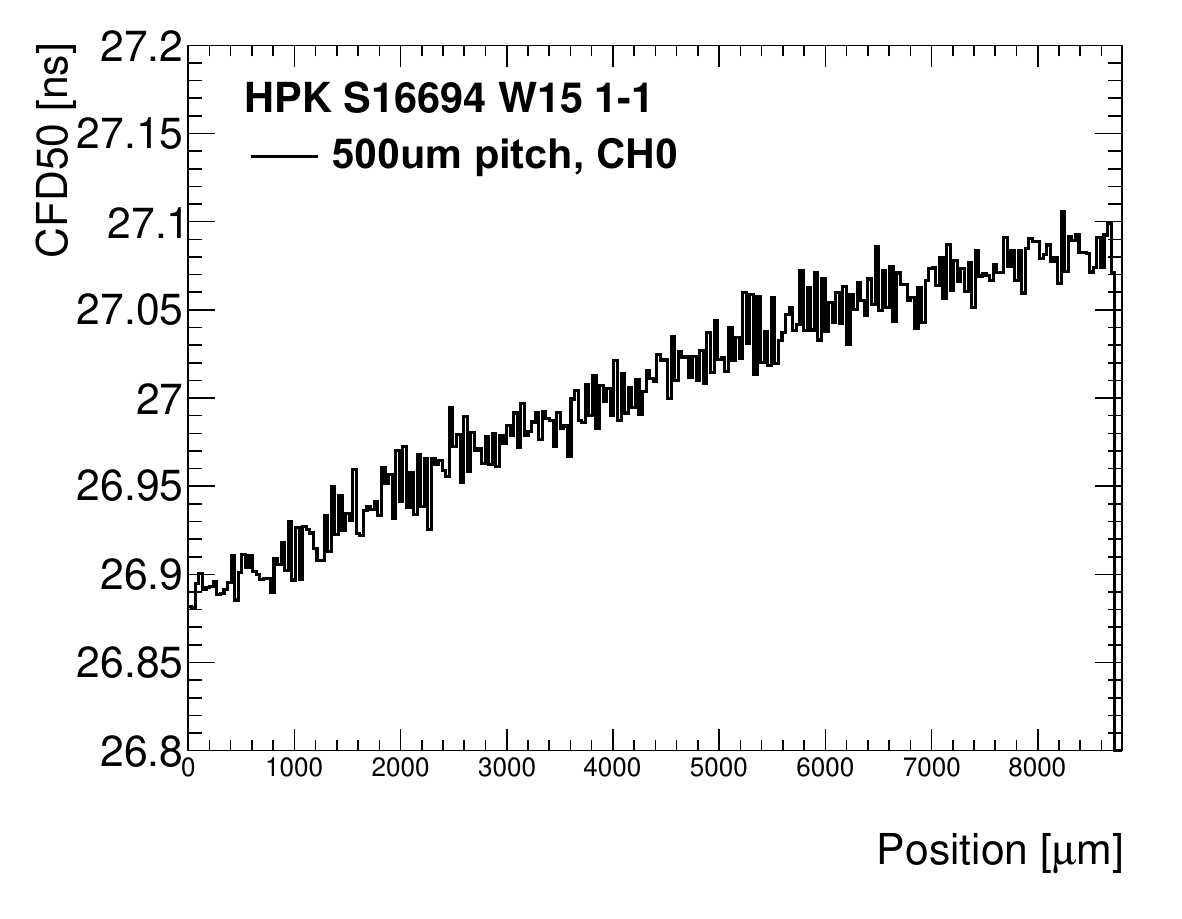}
    \caption{Left: time of 50\% of the Pmax vs laser position perpendicular to the strip. Right: time of 50\% of the Pmax vs laser position parallel to the strip, strip is connected on the left side.}
    \label{fig:strip_time}
\end{figure}

\begin{figure}[h]
    \centering
    \includegraphics[width=0.3\linewidth]{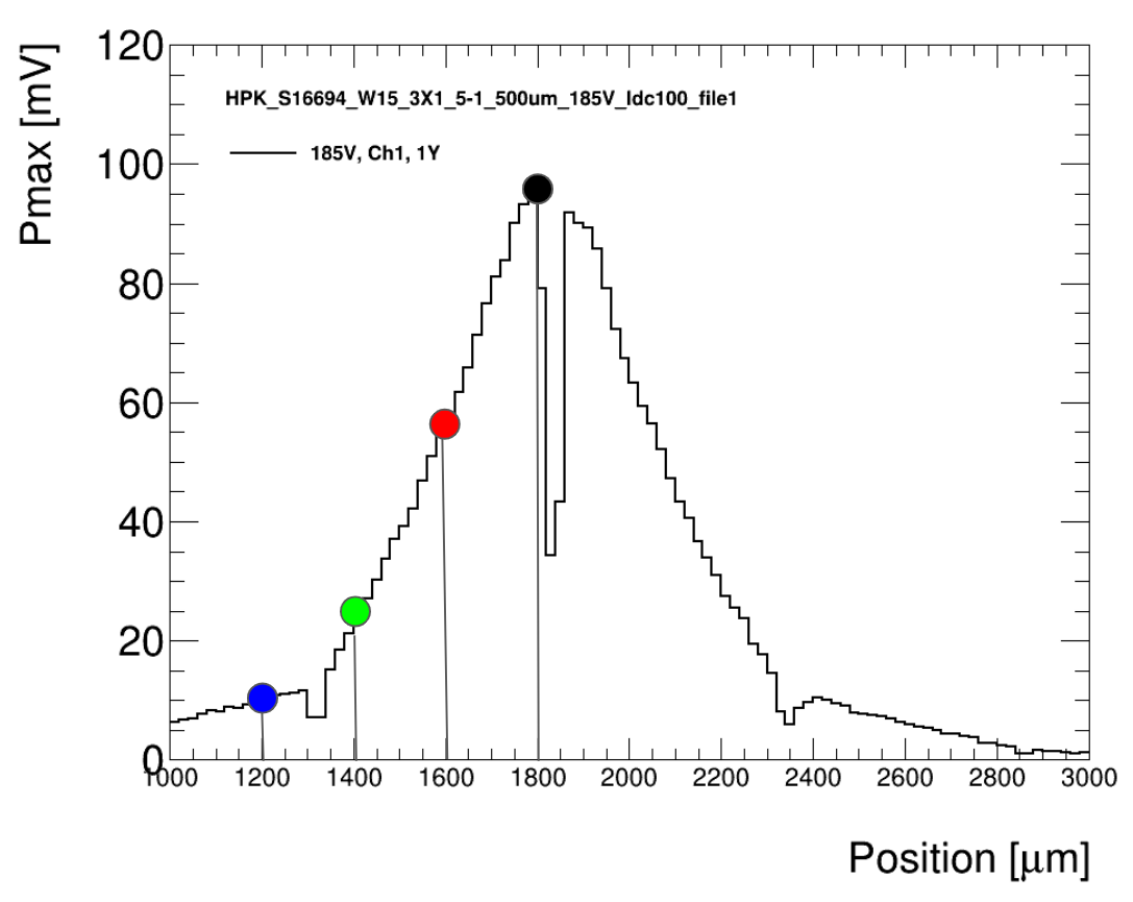}
    \includegraphics[width=0.3\linewidth]{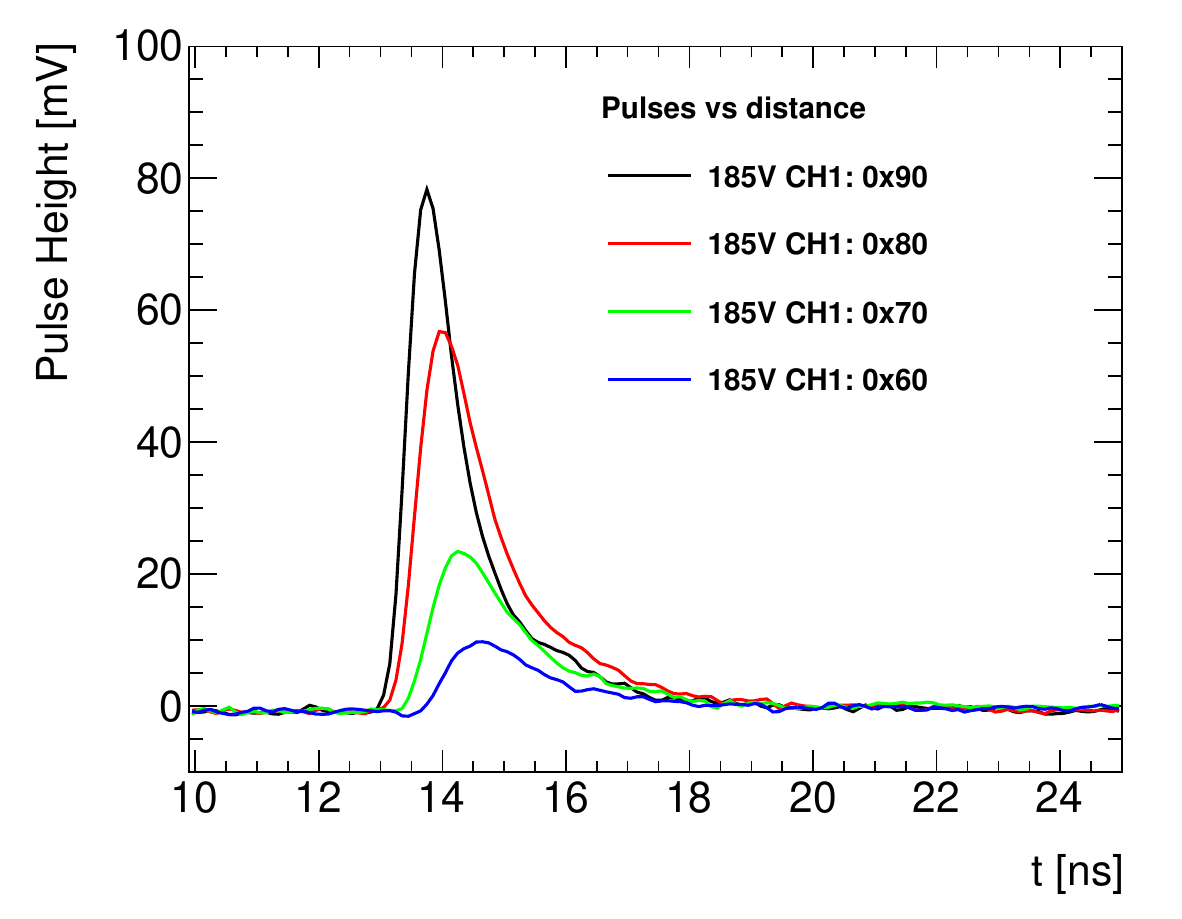}
    \includegraphics[width=0.3\linewidth]{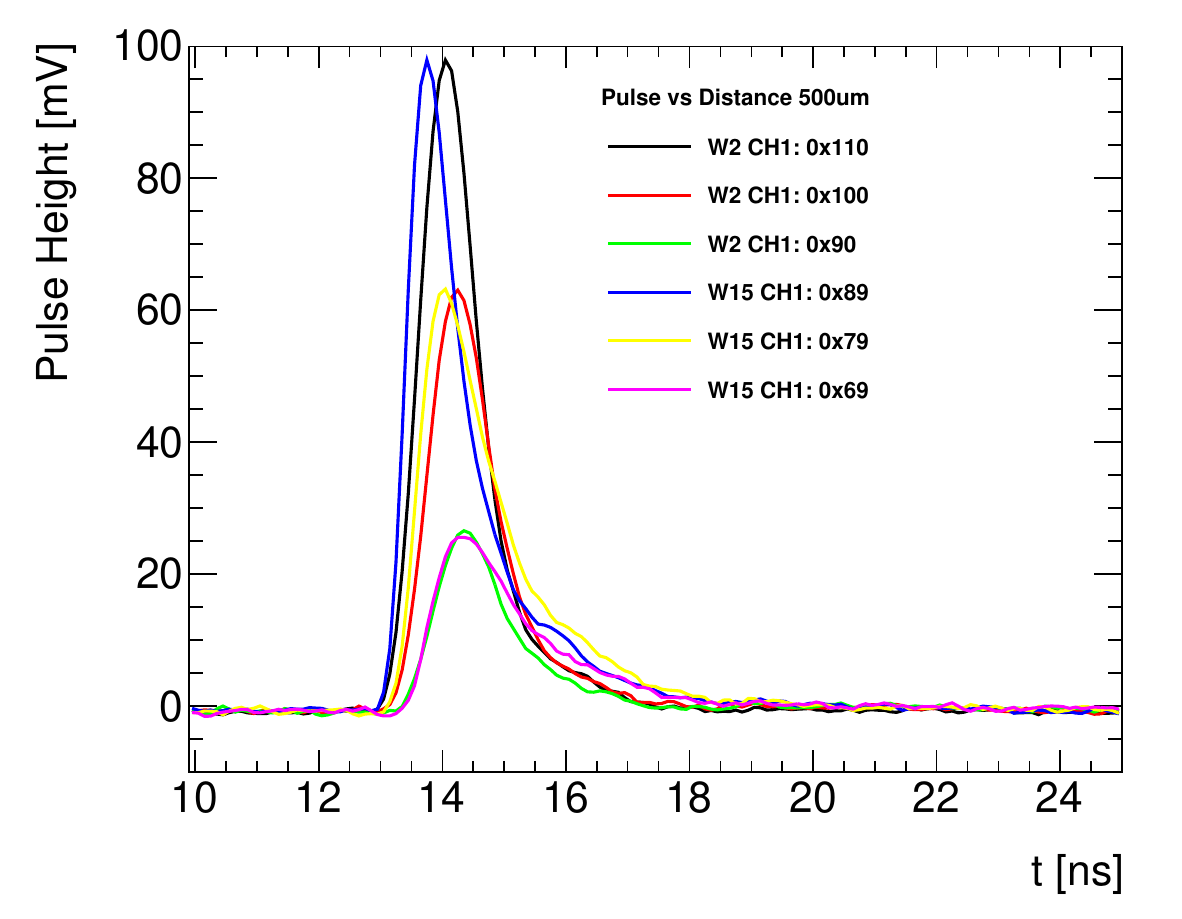}
    \caption{Left: Pmax vs laser position perpendicular to the strip. Center: Waveform of the 50~$\mu$m-thick, 500~$\mu$m-pitch strip sensor for positions highlighted in (Left). Right: Waveform for three different positions of the 50~$\mu$m-thick and 30~$\mu$m-thick strip sensors, showing the difference in rise time.}
    \label{fig:strip_pulse}
\end{figure}

The variables derived from the waveforms are as follows: pulse maximum or Pmax is the maximum of the waveform measured in mV. Rise time ($R_T$) is the 10-90\% rise time of the waveform. CFD50 is the time of arrival at 50\% of the pulse Pmax to remove time walk uncertainty.
The Pmax as a function of laser injection position perpendicular to the strip for 30~$\mu$m thickness is shown in Fig.~\ref{fig:strip_pmax} (Left). The profile is similar for the three pitches next to the strip, but deviates after a few hundred $\mu$m.
The rise time is shown in Fig.~\ref{fig:strip_pmax} (Right) for 30~$\mu$m thickness, showing a similar behavior for the three pitches.
The CFD50 time of arrival for 30~$\mu$m thickness is shown in Fig.~\ref{fig:strip_time} (Left) for the direction perpendicular to the strip (signal propagation in the resistive N+) and in Fig.~\ref{fig:strip_time} (Right) for the direction parallel to the strip (signal propagation in metal). The signal propagation in the N+ for strips was measured to be around 1.25~$\mu$m/ps and independent of the pitch, while the signal propagation on the metal strip was measured to be around 40~$\mu$m/ps.
Waveforms as a function of position on the sensors are shown in Fig.~\ref{fig:strip_pulse}.

\subsection{Device performance}
\begin{figure}[h]
    \centering
    \includegraphics[width=0.9\linewidth]{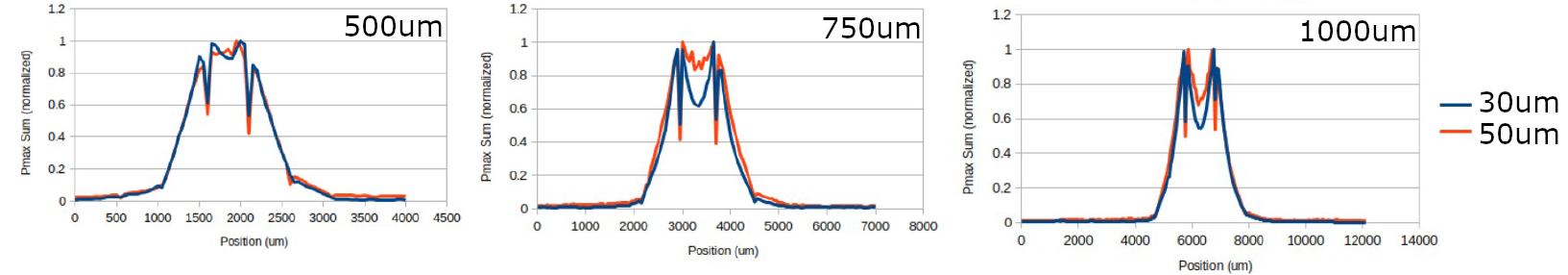}
    \caption{Normalized S/N of the sum of two neighboring strips for the three different pitches.}
    \label{fig:strip_SN}
\end{figure}

\begin{figure}[h]
    \centering
    \includegraphics[width=0.9\linewidth]{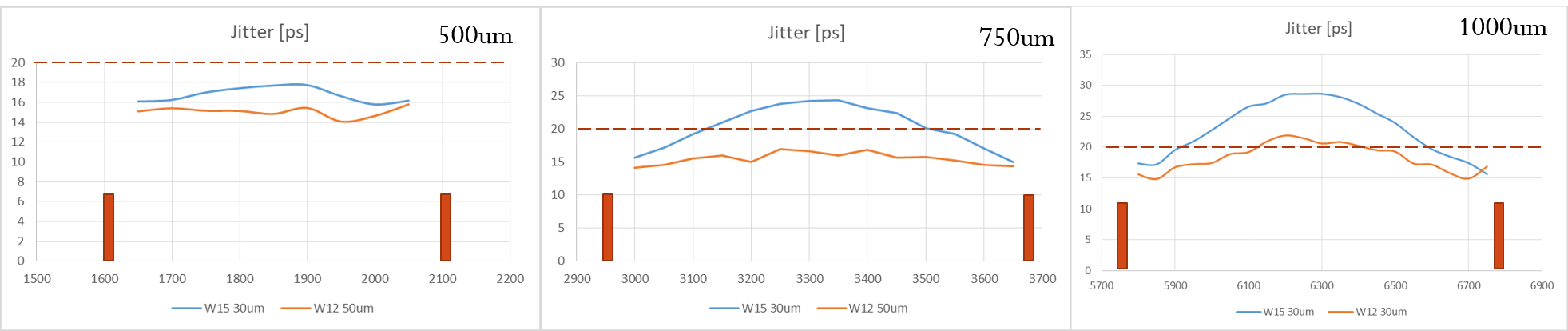}
    \caption{Calculated jitter component of the time resolution for the three different pitches. The dashed line represents a Jitter of 20~ps. The orange squares indicate the strips' position.}
    \label{fig:strip_jitter}
\end{figure}

\begin{figure}[h]
    \centering
    \includegraphics[width=0.9\linewidth]{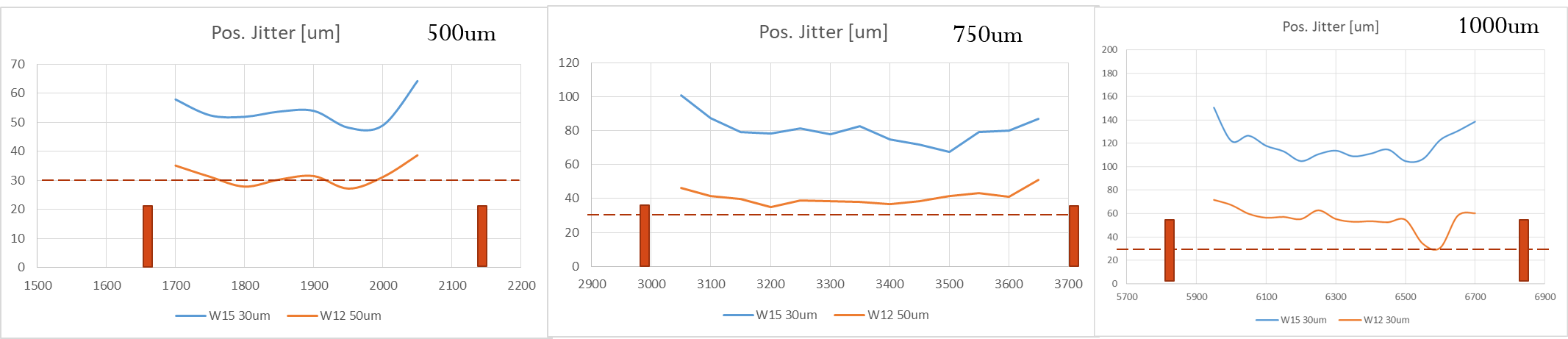}
    \caption{Calculated position jitter for the three different pitches. The dashed line represents a position jitter of 30~$\mu$m. The orange squares indicate the strips' position.}
    \label{fig:strip_pos_jitter}
\end{figure}

The performance of the device was evaluated following the method in~\cite{BISHOP2024169478}.
In short, the jitter component of the time resolution is calculated as $\mathrm{\sigma_{T,J}=R_T/(S/N)}$ and the position jitter is calculated as $\mathrm{\sigma_J(pos)=\sqrt{2}\left(\frac{dPos}{dFrac}\right)\frac{1}{S/N}}$ using the inverse of the Fractional slope (dFrac/dPos).
The pulse maximum was normalized to be around the Landau most probable value (MPV) of 1 minimum ionizing particle (MIP) following preliminary test beam results to around 60~mV under the strip for the 500~$\mu$m pitch, 50~$\mu$m-thick sensor. The other geometries are normalized with the same factor, as all measurements were executed on the same day and board. The 30~$\mu$m-thick sensor was normalized proportionally.
The normalized signal-to-noise for the three geometries and thicknesses is in Fig.~\ref{fig:strip_SN}, showing a significant signal loss for large pitches, in particular for the 30~$\mu$m-thick sensor.
$\sigma_{T,J}$ is shown in Fig.~\ref{fig:strip_jitter} and $\sigma_J(pos)$ is shown in Fig.~\ref{fig:strip_pos_jitter}.
The best performance is for the 500~$\mu$m pitch, 50~$\mu$m-thick sensor, which has a 20~ps jitter and a 30~$\mu$m positional jitter; all the other geometries and thicknesses show worse behavior.

\FloatBarrier

\section{Conclusions}
Strip and pixel sensors from the first full-size ePIC time-of-flight layer were characterized in the laboratory. 
The yield was estimated to be around 80\% for strips and 90\% for pixels. The cause of the low yield (57\%) for 50~$\mu$m thick strips was understood by the vendor and corrected in the following production.
The wafer characteristics were tested with the probe station, showing low variation across wafers of the same type of gain layer doping ($<$~0.5\%) and N+ resistivity ($<$~20\%) variation.
The strip sensors were tested with laser TCT and showed a loss of S/N for increased pitch, affecting the performance of the devices.
For the 50~$\mu$m thick, 500~$\mu$m pitch and 50~$\mu$m strip width sensors, the estimated Jitter component of the time resolution is under 20~ps and the position jitter is around 30~$\mu$m. All the other variations showed worse performance.
The signal propagation is 1.25~$\mu$m/ps in the N+ and 40~$\mu$m/ps in the strip metal.

\section{Acknowledgments}
We thank the technicians and students at SCIPP for the support and the time spent in the lab.
We acknowledge the collaboration with the KEK group (K. Nakamura et al.), the FNAL group (A. Apresyan et al.), the BNL group (A. Tricoli et al.), and the LBNL group (Z. Ye et al.). 
The sensors were produced with funds from the ePIC eRD112 effort for AC-LGADs development for the ePIC detector TOF layer.
We thank Ojas Khandelwal for the help in extracting the gain layer homogeneity.
This work was supported by the United States Department of Energy grant DE-SC0010107.

\bibliography{bib/TechnicalProposal,bib/hpk_fbk_paper,bib/HGTD_TDR,bib/SHIN,bib/nizam,bib/others}

@article{Stage:2025jre,
    author = "Stage, G. and others",
    title = "{Performance of neutron and proton irradiated AC-LGAD sensors}",
    eprint = "2503.16658",
    archivePrefix = "arXiv",
    primaryClass = "physics.ins-det",
    doi = "10.1016/j.nima.2025.171012",
    journal = "Nucl. Instrum. Meth. A",
    volume = "1082",
    pages = "171012",
    year = "2026"
}

@article{BISHOP2024169478,
title = {Long-distance signal propagation in {AC-LGAD}},
journal = {NIM A},
volume = {1064},
pages = {169478},
year = {2024},
issn = {0168-9002},
doi = {https://doi.org/10.1016/j.nima.2024.169478},
author = {C. Bishop et al.},
}

@article{MENZIO2024169526,
title = {First test beam measurement of the {4D} resolution of an {RSD} pixel matrix connected to a {FAST2} {ASIC}},
journal = {NIM A},
volume = {1065},
pages = {169526},
year = {2024},
issn = {0168-9002},
doi = {https://doi.org/10.1016/j.nima.2024.169526},
author = {L. Menzio et al.},
}

@article{Pellegrini:2014lki,
    author = "Pellegrini, G. and others",
    editor = "Unno, Yoshinobu and Fukazawa, Yasushi and Hou, Suen and Ohsugi, Takashi and Sadrozinski, Hartmut F. -W.",
    title = "{Technology developments and first measurements of Low Gain Avalanche Detectors (LGAD) for high energy physics applications}",
    doi = "10.1016/j.nima.2014.06.008",
    journal = "Nucl. Instrum. Meth. A",
    volume = "765",
    pages = "12--16",
    year = "2014"
}

@article{Sadrozinski:2013nja,
    author = "Sadrozinski, H. F. -W. and others",
    editor = "Pace, Emanuele and Talamonti, Cinzia and Bruzzi, Mara",
    title = "{Ultra-fast silicon detectors}",
    doi = "10.1016/j.nima.2013.06.033",
    journal = "Nucl. Instrum. Meth. A",
    volume = "730",
    pages = "226--231",
    year = "2013"
}

@article{Mazza:2019dkn,
    author = "Mazza, S. M.",
    collaboration = "ATLAS",
    title = "{A High-Granularity Timing Detector (HGTD) for the Phase-II upgrade of the ATLAS detector}",
    reportNumber = "ATL-LARG-PROC-2019-001",
    doi = "10.1088/1748-0221/14/10/C10028",
    journal = "JINST",
    volume = "14",
    number = "10",
    pages = "C10028",
    year = "2019"
}

@article{Ferrero:2022ynt,
    author = "Ferrero, M.",
    collaboration = "CMS",
    title = "{The CMS MTD Endcap Timing Layer: Precision timing with Low Gain Avalanche Diodes}",
    reportNumber = "CMS-CR-2021-249",
    doi = "10.1016/j.nima.2022.166627",
    journal = "Nucl. Instrum. Meth. A",
    volume = "1032",
    pages = "166627",
    year = "2022"
}

@article{Mandurrino,
author = {M. Mandurrino et al},
year = {2019},
month = {09},
pages = {1-1},
title = {Demonstration of 200, 100, and 50 $\mu$m pitch Resistive {AC}-Coupled Silicon Detectors ({RSD}) with 100\% fill-factor for {4D} particle tracking},
volume = {PP},
journal = {IEEE Electron Device Letters},
doi = {10.1109/LED.2019.2943242}
}

@misc{Hamamatsu,
  author = {HPK},
  url = {https://www.hamamatsu.com/jp/en.html},
 }

@article{AbdulKhalek:2021gbh,
    author = "Abdul Khalek, R. and others",
    title = "{Science Requirements and Detector Concepts for the Electron-Ion Collider}: {EIC Yellow Report}",
    eprint = "2103.05419",
    archivePrefix = "arXiv",
    primaryClass = "physics.ins-det",
    reportNumber = "BNL-220990-2021-FORE, JLAB-PHY-21-3198, LA-UR-21-20953",
    doi = "10.1016/j.nuclphysa.2022.122447",
    journal = "Nucl. Phys. A",
    volume = "1026",
    pages = "122447",
    year = "2022"
}

@article{Ott:2022itj,
    author = "J. Ott et al.",
    title = "{Investigation of signal characteristics and charge sharing in AC-LGADs with laser and test beam measurements}",
    reportNumber = "FERMILAB-PUB-22-839-PPD-QIS, BNL-223886-2023-JAAM",
    doi = "10.1016/j.nima.2022.167541",
    journal = "Nucl. Instrum. Meth. A",
    volume = "1045",
    pages = "167541",
    year = "2023"
}

@article{Heller:2022aug,
    author = "R. Heller et al.",
    title = "{Characterization of BNL and HPK AC-LGAD sensors with a 120 GeV proton beam}",
    eprint = "2201.07772",
    archivePrefix = "arXiv",
    primaryClass = "physics.ins-det",
    reportNumber = "FERMILAB-PUB-22-059-PPD",
    doi = "10.1088/1748-0221/17/05/P05001",
    journal = "JINST",
    volume = "17",
    number = "05",
    pages = "P05001",
    year = "2022"
}

\end{document}